\documentclass[aps,prd,showpacs,preprintnumbers,nofootinbib,superscriptaddress]{revtex4}
\usepackage[dvips]{graphicx}
\usepackage{bm,latexsym,amsmath,amssymb,amsfonts}
\newcommand*{\D}{{\rm d}}

\begin{document}

\title{Evolution of linear cosmological perturbations and its observational implications
in Galileon-type modified gravity}

\author{Tsutomu~Kobayashi}
\email[Email: ]{tsutomu"at"gravity.phys.waseda.ac.jp}
\affiliation{Department of Physics, Waseda University, Okubo 3-4-1, Shinjuku, Tokyo 169-8555, Japan}

\author{Hiroyuki~Tashiro}
\email[Email: ]{hiroyuki.tashiro"at"uclouvain.be}
\affiliation{Center for Particle Physics and Phenomenology (CP3), Universite catholique de Louvain,%
Chemin du Cyclotron, 2, B-1348 Louvain-la-Neuve, Belgium}

\author{Daichi~Suzuki}
\email[Email: ]{suzuki"at"gravity.phys.waseda.ac.jp}
\affiliation{Department of Physics, Waseda University, Okubo 3-4-1, Shinjuku, Tokyo 169-8555, Japan}

\begin{abstract}
A scalar-tensor theory of gravity can be made not only to account for the current cosmic acceleration,
but also to satisfy solar-system and laboratory constraints,
by introducing a non-linear derivative interaction for the scalar field.
Such an additional scalar degree of
freedom is called ``Galileon.''
The basic idea is inspired by the DGP braneworld, but
one can construct a ghost-free model that admits a self-accelerating solution.
We perform a fully relativistic analysis of linear perturbations in Galileon cosmology.
Although the Galileon model can mimic the background evolution of
standard $\Lambda$CDM cosmology, the behavior of perturbation is quite different.
It is shown that there exists a super-horizon growing mode in the metric and Galileon
perturbations at early times,
suggesting that the background is unstable.
A fine-tuning of the initial condition for the
Galileon fluctuation is thus required in order
to promote a desirable evolution of perturbations at early times.
Assuming the safe initial condition, we
then compute the late-time evolution of perturbations and
discuss observational implications in Galileon cosmology.
In particular, we find anticorrelations in the cross-correlation of the
integrated Sachs-Wolfe effect and large scale structure, similar to the normal branch of the DGP model.
\end{abstract}

\preprint{WU-AP/305/09}
\maketitle

\section{Introduction}

The origin of the current accelerated expansion of the universe~\cite{acc} is
one of the biggest mysteries in modern cosmology.
The conventional explanation is that it is caused by
a cosmological constant or dark energy, but then this idea indicates that
the universe today is mostly filled with some unknown energy-momentum component.
An alternative idea which has been explored actively in recent years
is that general relativity (GR) fails to hold at cosmological distances,
though GR is definitely the most successful theory of gravitation at shorter distances.

Typical models of modified gravity possess an extra propagating scalar degree of freedom,
and the prototype Lagrangian is often described by the Brans-Dicke theory~\cite{BD, Fujii}. 
However,
the theory in its simplest form clearly contradicts solar-system and laboratory experiments
except in the uninteresting parameter range where it is essentially indistinguishable from GR~\cite{Will}.
In more elaborate models, the extra scalar degree of freedom is carefully designed
so that the standard $\Lambda$-Cold Dark Matter ($\Lambda$CDM) model is mimicked
while solar-system and laboratory tests are safely evaded.
In some of $f(R)$ gravity models~\cite{fR} and their equivalent description in terms of
the ``chameleon'' scalar-tensor theory~\cite{Cham}, this is made possible by utilizing
a matter-dependent effective potential to hide the scalar field at short distances.
Another interesting way of suppressing the scalar field on short scales
is found in the Dvali-Gabadadze-Porrati (DGP) braneworld model~\cite{DGP},
in which a non-linear derivative self-coupling of the scalar is involved~\cite{decouple}.
Below a certain distance from the source on the DGP brane,
the non-linear effect of the brane bending (corresponding to the extra scalar mode)
plays an important role even for weak gravitational fields,
successfully screening the scalar-mediated force~\cite{Tanaka, nl-cosmo}.
Recently, a general class of scalar-tensor theories has been explored
that includes non-linear derivative interaction terms in the Lagrangian
and shares the same properties as the DGP effective theory~\cite{G1}. That is,
the theory leads to Lorentz invariant equations of motion having only second derivative
of the scalar field on a flat background.
Such a scalar field is dubbed a Galileon.
The Galileon theory has been covariantized and considered in curved backgrounds in Ref.~\cite{G2}.
Cosmology in Galileon modified gravity has been first studied in Ref.~\cite{G3}.
It is known that the self-accelerating solution in the DGP model suffers from
the presence of a ghost, and therefore it is interesting to investigate the ghost issue
also in the Galileon-type scalar-tensor theories.
In Ref.~\cite{SK}, a ghost-free Galileon model that admits a self-accelerating solution has been constructed.

In this paper, the evolution of cosmological perturbations in Galileon modified gravity is studied.
We develop a full cosmological perturbation theory at linear order,
with which we
can discuss the stability of the model and
compute observational signatures such as the integrated Sachs-Wolfe (ISW) effect and weak lensing.
These observations help us to distinguish between different dark energy/modified gravity models.
Though our analysis is not complete in the sense that the important non-linear effect is missed,
the linear theory is still valid on sufficiently large scales which we will be interested in.

The theory of gravitation we consider is described by the action~\cite{G1,G2,G3,SK}
\begin{eqnarray}
S =\int\D^4x\sqrt{-g}\left[
\phi R-\frac{\omega}{\phi}(\nabla\phi)^2+f(\phi)\Box\phi(\nabla\phi)^2+{\cal L}_{\rm m}
\right],
\end{eqnarray}
where $(\nabla\phi)^2:=g^{\mu\nu}\nabla_\mu\phi\nabla_\nu\phi$,
$\Box\phi:=g^{\mu\nu}\nabla_\mu\nabla_\nu\phi$, and ${\cal L}_m$ is the matter Lagrangian.
The matter Lagrangian does not depend on the Galileon field $\phi$.
The most general form of the Galileon action has been derived in Ref.~\cite{G2},
but in this paper we focus on the simple case which includes only $\Box\phi(\nabla\phi)^2$.
Variation with respect to the metric gives
the Einstein equations
\begin{eqnarray}
G_{\mu\nu}&=&\frac{T_{\mu\nu}}{2\phi}+
\frac{1}{\phi}\left(\nabla_\mu\nabla_\nu\phi-g_{\mu\nu}\Box\phi\right)
+\frac{\omega}{\phi^2}\left[\nabla_\mu\phi\nabla_\nu\phi-\frac{1}{2}g_{\mu\nu}
\left(\nabla\phi\right)^2\right]
\nonumber\\
&&\quad
-\frac{1}{\phi}\left\{
\frac{1}{2}g_{\mu\nu}\nabla_\lambda\left[f(\phi)\left(\nabla\phi\right)^2\right]
\nabla^\lambda\phi -\nabla_{(\mu}\left[f(\phi)\left(\nabla\phi\right)^2\right]
\nabla_{\nu)}\phi+f(\phi)\nabla_\mu\phi\nabla_\nu\phi\Box\phi
\right\},\label{basicEE}
\end{eqnarray}
and variation with respect to the Galileon field leads to the equation of motion
\begin{eqnarray}
&&R+\omega\left[\frac{2\Box\phi}{\phi}-\frac{\left(\nabla\phi\right)^2}{\phi^2}\right]
+2f(\phi)\left[\nabla_\mu\nabla_\nu\phi\nabla^\mu\nabla^\nu\phi-\left(
\Box\phi\right)^2+R_{\mu\nu}\nabla^\mu\phi\nabla^\nu\phi\right]
\nonumber\\&&\qquad\qquad
+\frac{\D^2f}{\D\phi^2}\left(\nabla\phi\right)^2\left(\nabla\phi\right)^2
+4\frac{\D f}{\D\phi}\nabla_\mu\phi\nabla_\nu\phi\nabla^{\mu}\nabla^\nu\phi=0.\label{basicGE}
\end{eqnarray}
The model that is most closely related to the DGP braneworld is given by $\omega=0$ and
$f(\phi)=r_c^2M_{\rm Pl}^2/\phi^3$, where $M_{\rm Pl}$ is the Planck scale and
$r_c$ is the crossover scale which is taken to be of the
order of the present Hubble scale to account for the late-time accelerated expansion of the universe~\cite{G3}.
In this paper we are mainly interested in the case with $f(\phi)=r_c^2/\phi^2$ and general $\omega$~\cite{SK},
but nevertheless the basic perturbation equations are provided without specifying the form of $f(\phi)$.

The paper is organized as follows.
In the next section,
we review the background evolution of Galileon cosmology and emphasize the presence of
the self-accelerating solution.
In Sec.~III the evolution of linear perturbations in Galileon cosmology is addressed
analytically and numerically.
Observational signatures in the Galileon model are discussed in Sec.~IV.
Finally, we draw our conclusions in Sec.~V.



\section{The background evolution}

We start with reviewing the background evolution of Galileon cosmology.
For the Friedmann-Robertson-Walker background metric,
$\D s^2=-\D t^2+a^2(t)\delta_{ij}\D x^i\D x^j$, the Einstein equations give
\begin{eqnarray}
3H^2&=&\frac{\rho}{2\phi}-3HI+\frac{\omega}{2} I^2+ \phi^2 f\left(3H-\frac{\alpha_1}{2}I\right)I^3,
\label{eq-Friedmann}
\\
-3H^2-2\dot H&=&\frac{p}{2\phi}+ \dot I+ I^2+2HI
+\frac{\omega}{2} I^2- \phi^2 f\left(\dot I+\frac{2+\alpha_1}{2}I^2\right)I^2,
\label{eq-Ray}
\end{eqnarray}
while the equation of motion for the Galileon field reduces to
\begin{eqnarray}
&&6\left(2H^2+\dot H\right)- \omega\left(2\dot I+I^2+6HI\right)
\nonumber\\&&\quad
-6\phi^2 f\left(2H\dot I+\dot HI+2HI^2+3H^2 I\right)I
+4\phi^2 f\alpha_1 I^2\dot I+ \phi^2 f\left(\alpha_1^2+3\alpha_1+\alpha_2\right)I^4=0,
\label{eq-phi}
\end{eqnarray}
where we defined $H(t)=\dot a/a$, $I(t):=\dot\phi/\phi$, and
$\alpha_n[\phi(t)]:=\D^n\ln f/\D\ln\phi^n$.
In the Jordan frame the energy-momentum conservation law remains the same as the standard one:
$\dot\rho+3H(\rho+p)=0$.


Let us consider a specific model with
\begin{eqnarray}
f(\phi)=\frac{r_c^2}{\phi^2},
\end{eqnarray}
for which a self-accelerating solution has been shown to exist~\cite{SK}.
At early times, $H^2\gg r_c^2$, we have an approximate solution for Eq.~(\ref{eq-phi}):
\begin{eqnarray}
I^2\simeq \frac{1}{r_c^2}\frac{2H^2+\dot H}{3H^2+\dot H}.
\end{eqnarray}
Since $I^2\sim r_c^{-2}\ll H^2$,
the Galileon field is constant on a cosmological time scale, $\phi\simeq 1/16\pi G$,
and the Einstein equations~(\ref{eq-Friedmann}) and~(\ref{eq-Ray})
reduce to the usual ones: $3H^2=8\pi G\rho$ and $3H^2+2\dot H=-8\pi G p$.
This is the cosmological version of the Vainshtein effect,
by which GR is recovered below a certain scale.
At late times we find an accelerating solution satisfying
\begin{eqnarray}
\frac{I}{H}\simeq s:=\frac{-1\pm\sqrt{-2-3\omega/2}}{1+\omega/2 },
\quad
H^2\simeq \frac{1}{r_c^2}\frac{s+2}{s^3}.
\end{eqnarray}
Note that in order for the self-accelerating solution to exist we require $4+3\omega <0$.
Interestingly, though $\phi$ is a ghost in this parameter range in the usual Brans-Dicke theory,
the Galileon theory does not suffer from ghost-like instabilities.

\begin{figure}[t]
  \begin{center}
    \includegraphics[keepaspectratio=true,height=70mm]{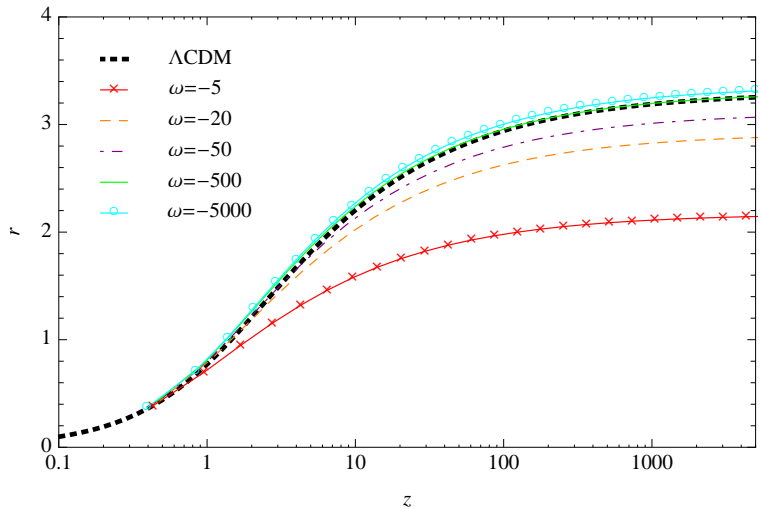}
  \end{center}
  \caption{Dimensionless physical distance for different $\omega$.}%
  \label{fig:bg.eps}
\end{figure}

It is easy to
solve numerically for the intermediate regime
between the early-time solution and the late-time accelerating one.
We can confirm that the early-time solution specified above indeed
evolves into the self-accelerating solution at late times, mimicking the
standard $\Lambda$CDM model~\cite{SK}.
In this $\Lambda$CDM-like solution, the parameter $r_c$ must be tuned so as to satisfy, say,
$\Omega_{\rm m}(a)[:=\rho/(6\phi H^2)] = 0.3$ at the present day.
In Fig.~\ref{fig:bg.eps} we show the dimensionless physical distance,
$r(z):=H_0\int^z_0 \D z'/H(z')$, for different $\omega$ as a function of redshift $z\,(:=1/a-1)$.
Except for small $|\omega|$ (say, $|\omega|\lesssim 10$),
the background evolution in Galileon cosmology is almost indistinguishable from the $\Lambda$CDM model. 
We shall see how this degeneracy is disentangled at perturbative order.

\section{Cosmological perturbations}

Let us study the linear evolution of scalar cosmological perturbations.
We work in the Newtonian gauge and consider a perturbed metric
\begin{eqnarray}
\D s^2 = -(1+2\Phi)\D t^2+a^2(1-2\Psi)\delta_{ij}\D x^i\D x^j.
\end{eqnarray}
The perturbation of the Galileon field is written as
\begin{eqnarray}
\delta\phi =\phi(t)\varphi(t, \Vec{x}),
\end{eqnarray}
while the perturbed energy-momentum tensor is given by
\begin{eqnarray}
\delta T_0^{\;0}=-\delta\rho,\quad\delta T_i^{\;0}=\partial_i\delta q,
\quad \delta T_i^{\;j} = \delta p\,\delta_i^{\;j}.
\end{eqnarray}
Here and hereafter an anisotropic stress perturbation is neglected.
Since the energy-momentum conservation holds in the Jordan frame, we have
\begin{eqnarray}
\dot{\delta\rho}+3H(\delta\rho+\delta p)-3\dot\Psi(\rho+ p)+
\frac{\nabla^2}{a^2}\delta q&=&0,
\label{cons-0}
\\
\dot{\delta q}+3H\delta q+(\rho+p)\Phi+\delta p &=&0,
\label{cons-i}
\end{eqnarray}
where $\nabla^2 = \delta^{ij}\partial_i\partial_j$.
The traceless part of the $(i, j)$ component of the Einstein equations reduces to 
\begin{eqnarray}
\Psi-\Phi = \varphi,
\end{eqnarray}
from which one can see that the Galileon field gives rise to
an effective anisotropic stress.
It is convenient to define
\begin{eqnarray}
\Omega := \frac{1}{2}(\Psi+\Phi),
\end{eqnarray}
because
weak lensing and the ISW effect are determined through
this combination of the metric potentials.
Using this variable one may write
$\Psi = \Omega+\varphi/2$ and $\Phi=\Omega-\varphi/2$.
The perturbed Einstein equations and the perturbed equation of motion for the Galileon field
are summarized in Appendix~A without specifying the function $f(\phi)$.
We will, however, focus
on the specific model with $f(\phi)=r_c^2/\phi^2$ in the bulk of the present paper.


\subsection{Early-time behavior}\label{E-t_b}

At early times the effect of the Galileon in the background can be neglected: $H^2\gg I^2\sim r_c^{-2}$.
Let us look at the behavior of cosmological perturbations in this regime.
Keeping leading order terms in the Einstein equations, we obtain
\begin{eqnarray}
6H\left(\dot \Omega+H\Omega\right)-2\frac{\nabla^2}{a^2}\Omega
+8\pi G \delta\rho&=&16\pi G\rho\varphi+
 r_c^2I^2\left(
-9H\dot\varphi+\frac{\nabla^2}{a^2}\varphi
\right),
\label{et00}
\\
-2\left(\dot \Omega+H\Omega\right)-8\pi G\delta q&=& r_c^2I^2\left(
\dot\varphi-3H\varphi\right),
\label{et0i}
\\
2\left[\ddot\Omega+4H\dot\Omega+\left(3H^2+2\dot H\right)\Omega\right]
-8\pi G\delta p&=&-16\pi Gp\varphi- r_c^2I^2\ddot\varphi,\label{etij}
\end{eqnarray}
while the equation of motion for the Galileon becomes a massless wave equation
\begin{eqnarray}
\ddot\varphi+\left(3H+\frac{\dot H}{H}\right)\dot\varphi-\frac{2}{3}
\frac{\nabla^2}{a^2}\varphi =0.
\label{et-phi}
\end{eqnarray}

It is clear
from Eq.~(\ref{et-phi}) that the fluctuation oscillates inside the sound horizon:
\begin{eqnarray}
\varphi\simeq \frac{{\cal C}_0}{aH^{1/2}}\cos\left(\sqrt{\frac{2}{3}}k\int^t\frac{\D t}{a}+\theta_0\right),
\end{eqnarray}
where we have moved to the Fourier space by promoting $\nabla^2$ to $-k^2$.
Here ${\cal C}_0$ and $\theta_0$ are the integration constants.

On super-horizon scales we ignore the spatial derivative term and obtain
\begin{eqnarray}
\varphi \simeq {\cal C}_1 +{\cal C}_2\int^t\!\!\frac{\D t}{a^3 H},
\end{eqnarray}
where ${\cal C}_1$ and ${\cal C}_2$ are the integration constants.
Suppose that the universe is dominated by a single fluid with the equation of state $w$.
It follows that
the second term grows like $\sim t^{2w/(1+w)}$ (respectively $\sim \ln t$)
for $w>0$ (respectively $w=0$).
This growing behavior is caused by the term $(\dot H/H)\dot\varphi$ in Eq.~(\ref{et-phi}), which
decreases the effect of the Hubble friction.
The growth of $\varphi$ then induces the growth of the metric potential
on super-horizon scales:
\begin{eqnarray}
\Omega \sim \int^t\!\!\frac{\D t}{a^3 H}.
\end{eqnarray}
The growth of the long wavelength modes suggests that the homogeneous background
we are considering is unstable.
The logarithmic growth in the matter-dominated era is rather mild, but
in the radiation-dominated stage the ${\cal C}_2$ mode grows like $\sim t^{1/2}$, which would be more dangerous.
Even if ${\cal C}_2=0$ initially so that the above growing solution is avoided, the ${\cal C}_1$ mode influences
the evolution of the comoving density perturbation defined by $\rho\Delta_c:=\delta\rho-3H\delta q$.
Indeed, combining Eqs.~(\ref{et00}) and~(\ref{et0i}) one obtains the generalized Poisson equation,
\begin{eqnarray}
\Delta_c = \frac{1}{3}\frac{\nabla^2}{a^2 H^2} \left(2\Omega+ r_c^2I^2\varphi\right)
-2r_c^2I^2\frac{\dot \varphi}{H}+ \left(2-3r_c^2I^2\right)\varphi,
\end{eqnarray}
and for long wavelength modes one sees that the last term governs the evolution of
the comoving density perturbation: $\Delta_c \simeq$ constant $\propto{\cal C}_1$.
This will drastically change the growth of the density perturbation, as will be seen
in a numerical example.

Since the evolution equation for $\varphi$ is decoupled from the Einstein equations,
we can have an early-time solution $\varphi = 0$ by tuning the initial conditions
so that $\varphi_i=0$ and $\dot\varphi_i=0$ at some $t=t_i$.
With this initial conditions,
the evolution of the metric and density perturbations is identical to
that in the $\Lambda$CDM model at early times, and the above unwanted behavior could thus be removed.
This seems to be the most safe choice of the initial conditions.
When
the effect of the Galileon becomes large in the background and
the universe enters the accelerating stage, the fluctuation $\varphi$
will be excited and consequently the evolution of the metric and density perturbations will be modified.
From now on, we will be focusing on this late-time effect.



\subsection{Quasi-static approximation}

For the subhorizon evolution of the density perturbation in the matter-dominated stage,
the quasi-static approximation is often employed in the literature, by which
the relevant equations can be much simplified.
In Galileon cosmology,
the closed evolution equation for the density perturbation
in the quasi-static approximation was already derived in Ref.~\cite{SK}.
Here, we rederive the equation just for completeness.
Later we will confirm the validity of the approximation on subhorizon scales
by numerical calculations.

Making the approximation ${\cal O}(\nabla^2\Phi/a^2)\gg {\cal O}(H^2\Phi)$ inside the Hubble horizon,
the conservation equations~(\ref{cons-0}) and~(\ref{cons-i}) reduce to
\begin{eqnarray}
\ddot{\Delta}_c+2H\dot{\Delta}_c\simeq \frac{\nabla^2}{a^2}\Phi.
\end{eqnarray}
(This equation holds exactly on all scales in the $\Lambda$CDM model thanks to the Einstein equations.)
Picking up the spatial derivative terms in the modified Einstein equations
and the Galileon equation of motion, we find
\begin{eqnarray}
 r_c^2I^2\frac{\nabla^2}{a^2}\varphi+
2\frac{\nabla^2}{a^2}\Omega- \frac{\rho\Delta_c}{2\phi}&\simeq&0,
\\
\left[\beta(t)+ r_c^2I^2\left(1+ r_c^2I^2\right)\right]\frac{\nabla^2}{a^2}\varphi
+2\left(1+ r_c^2I^2\right)\frac{\nabla^2}{a^2}\Omega&\simeq&0,\label{qs-2}
\end{eqnarray}
where $\beta(t):=3+2\omega + r_c^2\left(4\dot I+2I^2+8HI- r_c^2I^4\right)$.
Using these equations, we obtain the evolution equation for the comoving density perturbation,
\begin{eqnarray}
\ddot{\Delta}_c+2H\dot{\Delta}_c\simeq
\frac{\rho\Delta_c}{4\phi}\left[1+\frac{(1+ r_c^2I^2)^2}{\beta(t)}\right],
\end{eqnarray}
and
the modified Poisson equation,
\begin{eqnarray}
\frac{\nabla^2}{a^2}\Omega = \left[1+\frac{ r_c^2I^2(1+ r_c^2I^2)}{\beta(t)}\right]
\frac{\rho\Delta_c}{4\phi}.
\end{eqnarray}
Since $\phi\simeq(16\pi G)^{-1}$ and $\beta\sim H/I \gg 1$ at early times,
one can confirm that $\varphi\simeq 0$ and the $\Lambda$CDM result is recovered.
In other words,
the quasi-static approximation will no longer be valid
in the general early-time evolution with $\varphi\neq 0\neq\dot\varphi$.


We also have the Galileon fluctuation in terms of the density perturbation:
\begin{eqnarray}
\beta \frac{\nabla^2}{a^2}\varphi \simeq-(1+ r_c^2I^2)\frac{\rho\Delta_c}{2\phi}.
\label{linearphi}
\end{eqnarray}
Equation~(\ref{linearphi}) would have possible non-linear terms of order $r_c^2(\nabla^2\varphi/a^2)^2$.
Such terms can be neglected provided that $\nabla^2\varphi/a^2\lesssim \beta/r_c^2$.
Since $\Delta_c\sim(\beta/H^2)(\nabla^2\varphi/a^2)$, this condition is translated to
$\Delta_c\lesssim \beta^2/(r_c H)^2\sim 1$.
Non-linear terms such as $r_c^2\nabla^2\varphi/a^2(\nabla\varphi/a)^2$ and
$r_c^2(\nabla\varphi/a)^4$ are suppressed by a factor $\varphi \,(\ll 1)$.
Therefore, the linear approximation is valid as long as $\Delta_c\lesssim 1$.
The non-linear effect would suppress the deviation from GR via the Vainshtein mechanism.
Non-linear screening of the scalar degree of freedom is an important aspect of the Galileon-type theories,
but it is beyond the scope of the present paper.

\subsection{Numerical solutions}\label{num_calc}

\begin{figure}[t]
  \begin{center}
    \includegraphics[keepaspectratio=true,height=70mm]{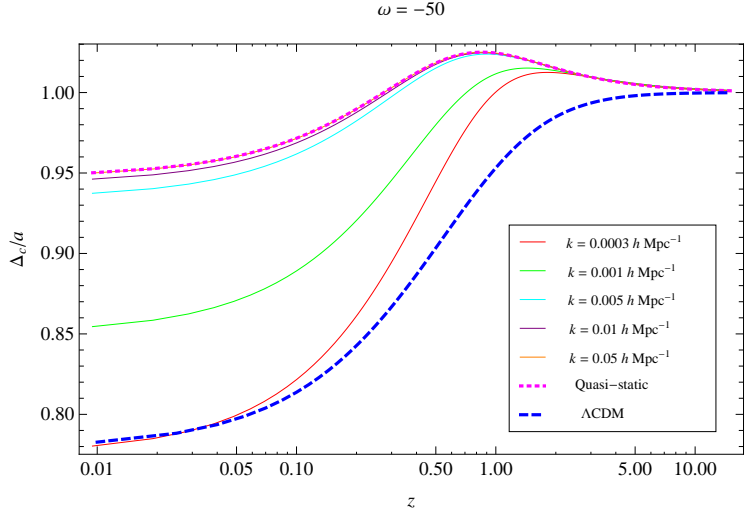}
  \end{center}
  \caption{Linear growth factor for different $k$.
  The initial condition for the numerical calculation is $\Omega_i=1$,
  $\dot\Omega_i=0$, and $\varphi_i=0=\dot\varphi_i$. The plotted value is divided by
  the initial value of the growth factor.
  The parameters are given by $\omega=-50$ and $\Omega_{{\rm m}0}=0.3$.
  Results obtained from the quasi-static approximation and in the $\Lambda$CDM model are also shown.}%
  \label{fig:D50.eps}
\end{figure}
\begin{figure}[t]
  \begin{center}
    \includegraphics[keepaspectratio=true,height=70mm]{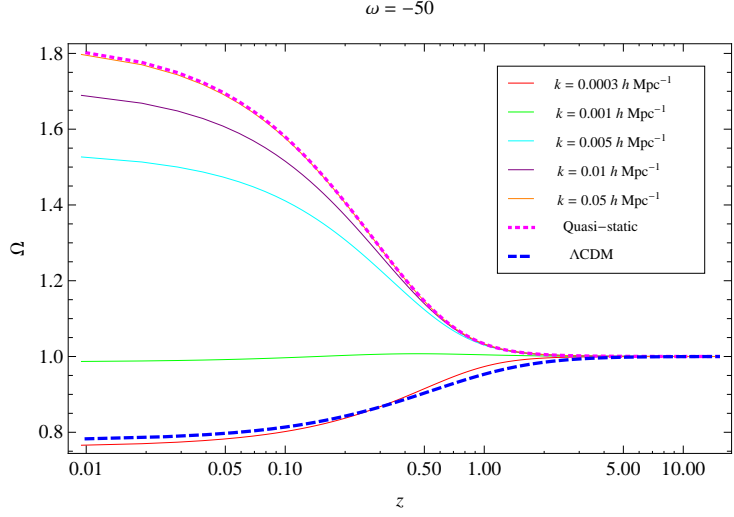}
  \end{center}
  \caption{Evolution of the potential $\Omega$ for different $k$.
  The same parameters as in Fig.~\ref{fig:D50.eps} are used.}%
  \label{fig:O50.eps}
\end{figure}
\begin{figure}[t]
  \begin{center}
    \includegraphics[keepaspectratio=true,height=70mm]{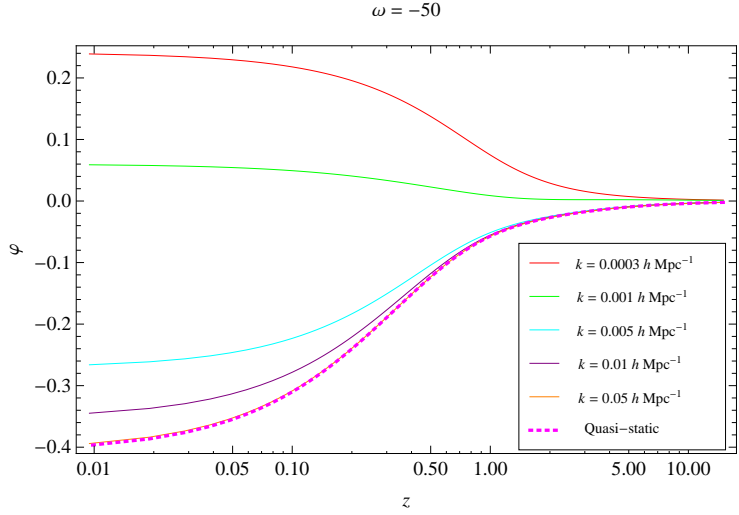}
  \end{center}
  \caption{Evolution of the Galileon fluctuation $\varphi$ for different $k$.
  The same parameters as in Fig.~\ref{fig:D50.eps} are used.}%
  \label{fig:P50.eps}
\end{figure}
\begin{figure}[t]
  \begin{center}
    \includegraphics[keepaspectratio=true,height=70mm]{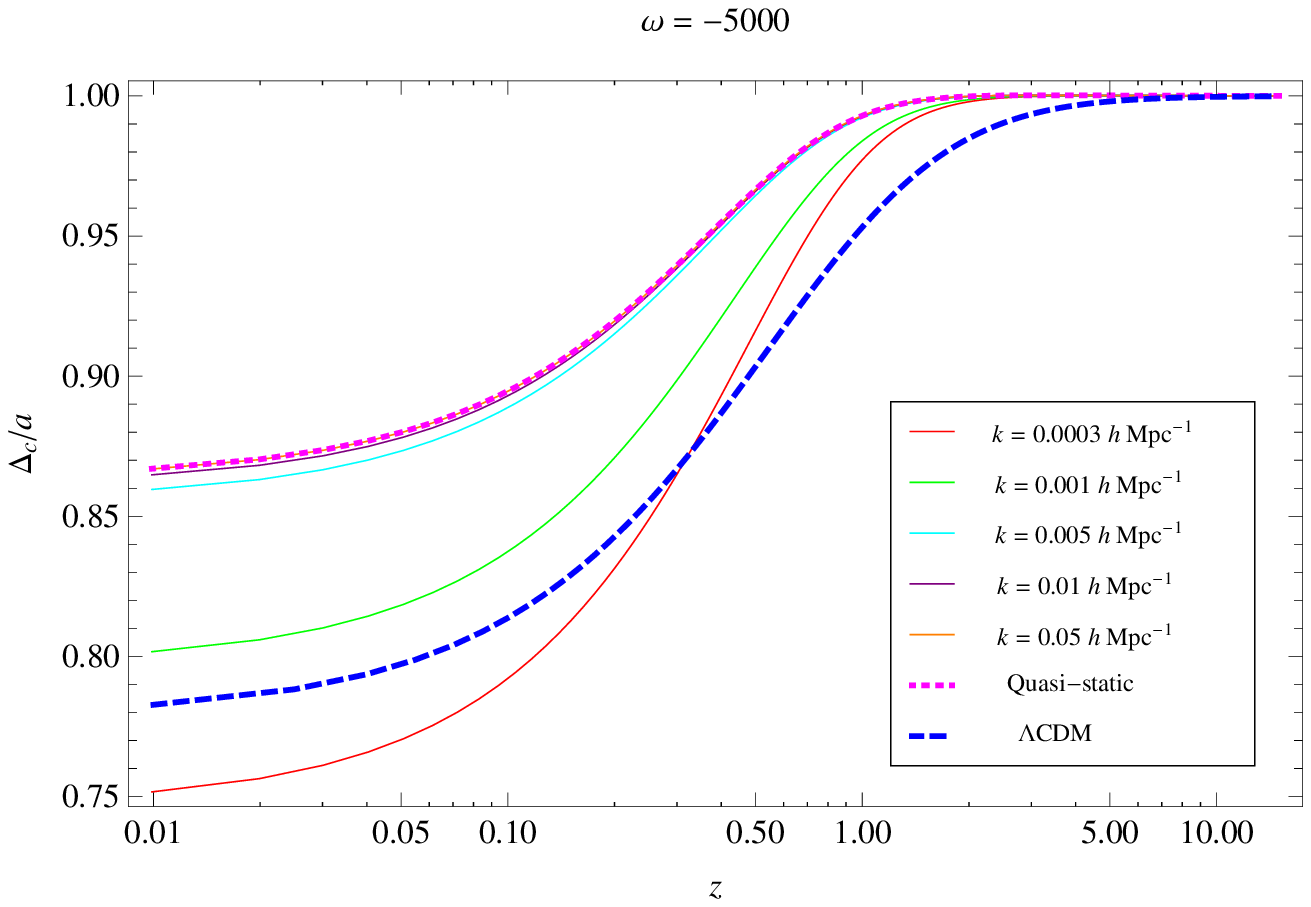}
  \end{center}
  \caption{Same as Fig.~\ref{fig:D50.eps}, but with $\omega =-5000$.}%
  \label{fig:D5000.eps}
\end{figure}
\begin{figure}[t]
  \begin{center}
    \includegraphics[keepaspectratio=true,height=70mm]{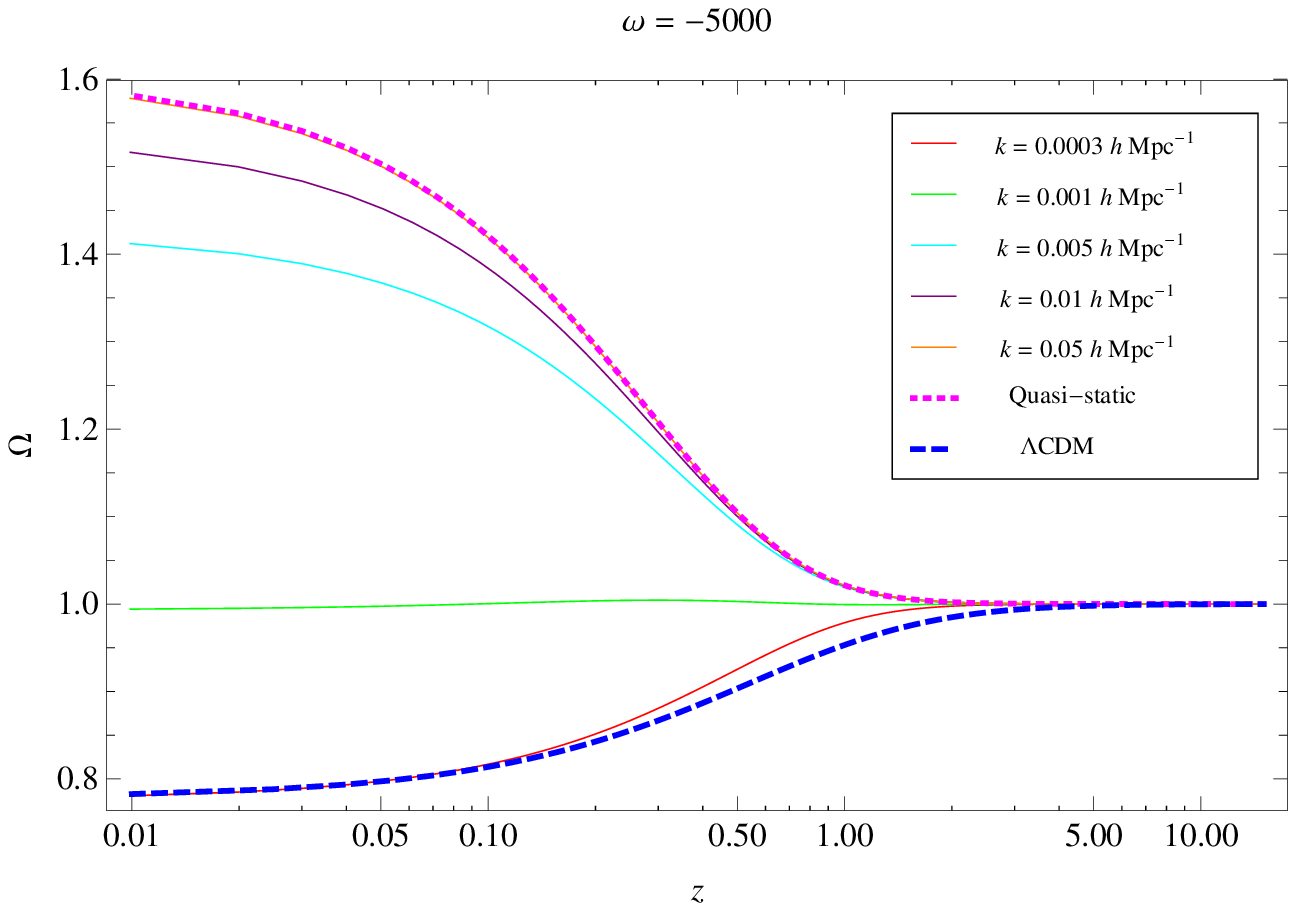}
  \end{center}
  \caption{Same as Fig.~\ref{fig:O50.eps}, but with $\omega=-5000$. }%
  \label{fig:O5000.eps}
\end{figure}
\begin{figure}[t]
  \begin{center}
    \includegraphics[keepaspectratio=true,height=70mm]{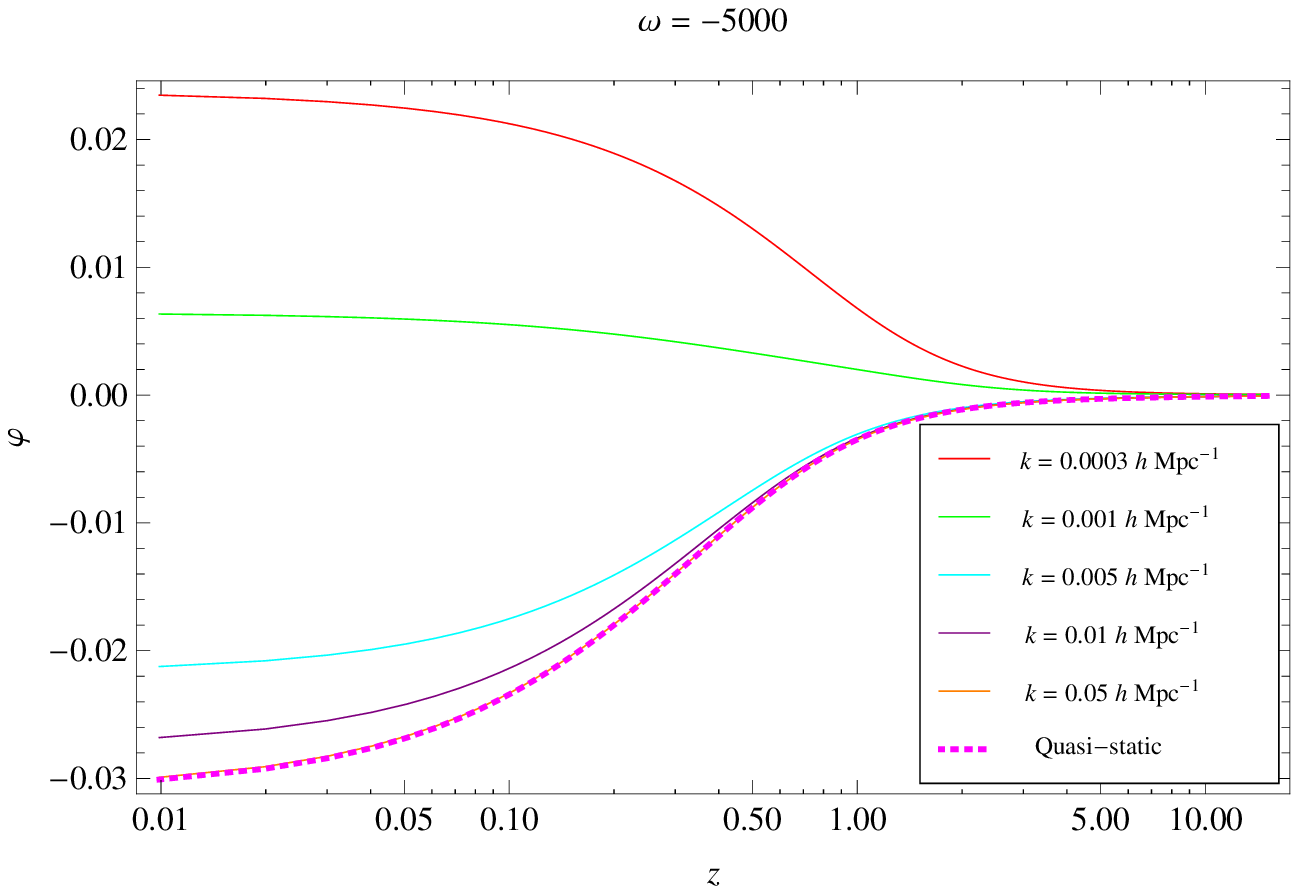}
  \end{center}
  \caption{Same as Fig.~\ref{fig:P50.eps}, but with $\omega=-5000$. }%
  \label{fig:P5000.eps}
\end{figure}

We are interested in the matter dominant era and the accelerating stage after that, so that
we neglect the pressure of matter: $p=0=\delta p$. Then,
the trace part of the Einstein equations and the equation of motion for the Galileon imply
\begin{eqnarray}
&&2\ddot\Omega+(8H+3I- r_c^2I^3)\dot\Omega- 2r_c^2I^2\dot I
\Omega
-\frac{3+2\omega}{2}I \dot\varphi 
+ r_c^2 I\left[I\ddot\varphi
+\left(2\dot I+\frac{I^2}{2}\right)\dot\varphi
+I\dot I\varphi
\right]=0
\end{eqnarray}
and
\begin{eqnarray}
&&(3+2\omega)\left[\ddot\varphi+3H\dot\varphi-\frac{\nabla^2}{a^2}\varphi\right]
- \left[6(2H^2+ \dot H)- \omega\left(2\dot I+I^2+6HI\right)\right]\left(2\Omega- \varphi \right)
+2\left[3\ddot\Omega+(15H-4\omega I)\dot\Omega-\frac{\nabla^2}{a^2}\Omega\right]
\nonumber\\&&\quad
=r_c^2\biggl\{
- (12H+5I)I\ddot\varphi
- \left(12H\dot I+12\dot HI+27HI^2+36H^2I+10I\dot I+6I^3\right)\dot\varphi
+ \left(4\dot I+3I^2+8HI\right)\frac{\nabla^2}{a^2}\varphi
\nonumber\\&&\qquad
+2I\left[3I\ddot\Omega+\left(6\dot I+10I^2+27HI\right)\dot\Omega +I\frac{\nabla^2}{a^2}\Omega\right]
\biggr\}.
\end{eqnarray}
Given the background evolution and the initial conditions for $\Omega$ and $\varphi$,
the above two equations can be solved numerically in the Fourier space.
The matter density perturbation on comoving slices,
$\Delta_c$,
is then computed by combining the $(0,0)$ and $(0,i)$ components of the Einstein equations as follows:
\begin{eqnarray}
-2\frac{\nabla^2}{a^2}\Omega+\frac{\rho\Delta_c}{2\phi}-\frac{\rho}{\phi}\varphi &=&
-3I\left[\dot\Omega+\left(H+\frac{I}{2}\right)\Omega\right]
-\frac{3+2\omega}{2}I\left( \dot\varphi+3H \varphi-I \Omega\right)
\nonumber\\&&
+ r_c^2I^2\left[
3I\dot\Omega+ (9H+4I)I\Omega-\left(6H+\frac{5}{2}I\right)\dot\varphi+\frac{\nabla^2}{a^2}\varphi
-\left(9H^2+\frac{9}{2}HI+ I^2\right)\varphi
\right].\label{genelarizedP}
\end{eqnarray}
This may be regarded as the generalization of the Poisson equation.

Numerical solutions of $\Delta_c$, $\Omega$, and $\varphi$
are shown in Figs.~\ref{fig:D50.eps}--\ref{fig:P5000.eps}
for $\omega=-50$ and $\omega=-5000$ as functions of redshift $z$.
For the present-day value of $\Omega_{\rm m}(a)$ we used $\Omega_{{\rm m}0}=0.3$.
In order for the perturbation evolution in the early stage to be the same as in the $\Lambda$CDM model,
the initial condition $\varphi_i=0=\dot\varphi_i$ has been used.
We find that the quasi-static approximation works well for the modes with $k>0.01\, h $Mpc$^{-1}$,
while the evolution deviates from the quasi-static solutions on larger scales.
As clearly seen, the behavior of the growth rate and metric potential differs from the $\Lambda$CDM case,
which will lead to different observational signatures.
In particular, it is interesting to note that
the metric potential $\Omega$ grows even when the growth rate decays.
The late-time growth of $\Omega$ is also found in the {\em normal (i.e., non-self-accelerating) branch}
of the DGP model~\cite{ckss}.

\begin{figure}[t]
  \begin{center}
    \includegraphics[keepaspectratio=true,height=100mm]{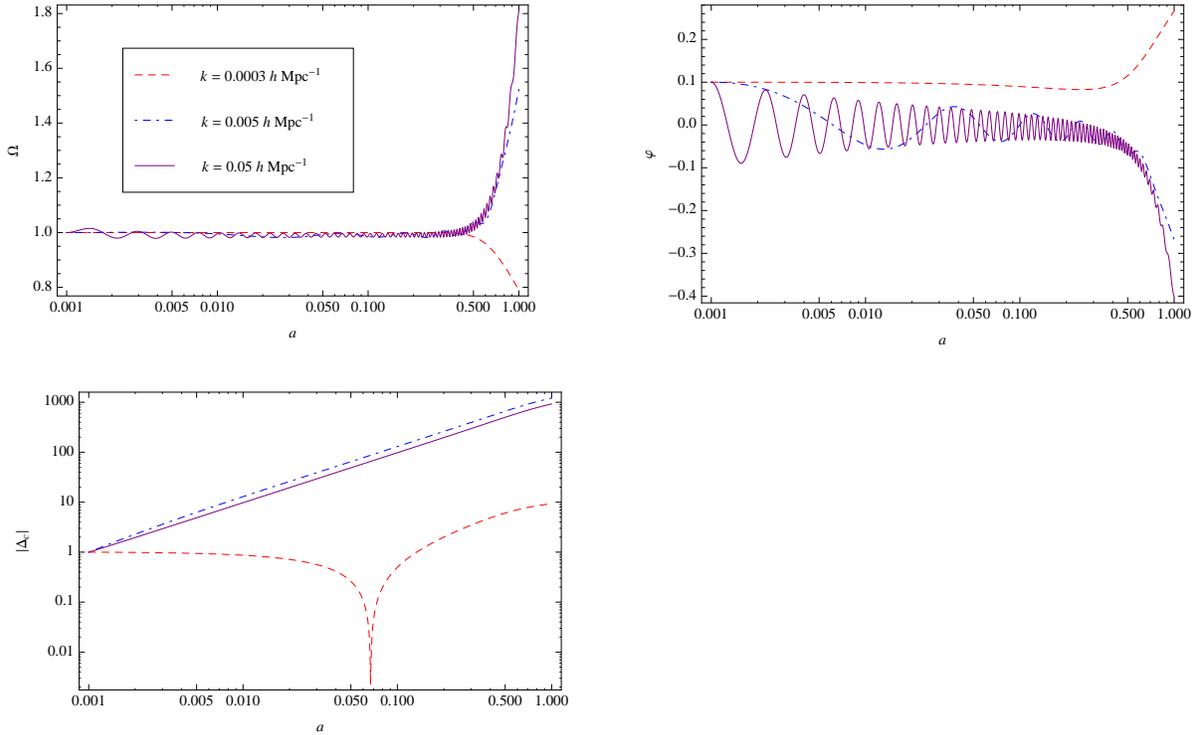}
  \end{center}
  \caption{Evolution of cosmological perturbations with the initial condition
  $\Omega_i=1$, $\varphi_i=0.1$, and $\dot\Omega_i=\dot\varphi_i=0$ at $a_i=10^{-3}$.
  The parameters are given by $\omega=-50$ and $\Omega_{{\rm m}0}=0.3$. The plotted value is $\Delta_c$
  divided by its initial value.}%
  \label{fig:initial.eps}
\end{figure}

To complete the discussion in Sec.~\ref{E-t_b},
we illustrate the effect of the different choices of the initial condition
just by showing the example obtained from the initial condition $\Omega_1=1$, $\dot\Omega_i=0$,
$\varphi_i=0.1\times\Omega_i$, and $\dot\varphi_i=0$ at $a_i=10^{-3}$ in Fig.~\ref{fig:initial.eps}.
One sees the oscillating behavior of $\Omega$ and $\varphi$ for short wavelength modes.
On large scales the comoving density perturbation does not grow, as explained earlier.

\section{Observational consequences}

We separate the evolution of the metric and density perturbations from the initial amplitude of $\Omega$ and write
$\Omega(k; \eta)={\cal D}_\Omega(k; \eta)\Omega(k; \eta_i)$ and
$\Delta_c(k; \eta)={\cal D}_\Delta(k; \eta)\Omega(k; \eta_i)$,
where $\eta$ is the conformal time, $\eta=\int\D t/a$,
and $\eta_i$ is some initial time.
The angular power spectrum between two observed fields $X$ and $Y$ on the sky is given 
in term of the angular transfer function $I_l^{X,Y}(k)$ by
\begin{eqnarray}
C_l^{XY} =4\pi  \int\frac{\D k}{k}
\left[\frac{9}{25}\Delta_{\cal R}^2(k) T^2(k)\right]I_l^X(k)I_l^Y(k).
\end{eqnarray}
Here, $\Delta^2_{\cal R}$ is the primordial spectrum of the comoving curvature perturbation, ${\cal R}$,
and during matter domination we have $\Omega = -(3/5){\cal R}$ because
the fine-tuned initial condition $\varphi_i=0=\dot\varphi_i$ implies that
the perturbation evolution before the last scattering ($\eta=\eta_*$) is identical to
the standard one in the $\Lambda$CDM model. The transfer function $T(k)$ is to be defined shortly.
We consider weak lensing, the ISW effect, and large scale structure (LSS) in the following discussion.
For weak lensing we have
\begin{eqnarray}
I_l^{\rm WL}(k)=\int^{\eta_0}_{\eta_{*}}\D \eta\,\frac{ W(\eta)}{(\eta_0-\eta)^2}{\cal D}_\Omega(k;\eta)
j_l[k(\eta_0-\eta)],\label{IWL}
\end{eqnarray}
where $j_i$ is the spherical Bessel function, $\eta_0$ is the present time, and the window function is defined as
\begin{eqnarray}
W(\eta):=2(\eta_0-\eta)\int^\eta_0\D\eta'\frac{\eta-\eta'}{\eta_0-\eta'}n(\eta'),
\end{eqnarray}
with $n(\eta)$ being a normalized distribution of the source.
Similarly, for the ISW effect $I_l^{\rm ISW}$ is given by
\begin{eqnarray}
I_l^{\rm ISW}(k)=2\int^{\eta_0}_{\eta_{*}}\D \eta\,\partial_\eta{\cal D}_\Omega(k;\eta)
j_l[k(\eta_0-\eta)].\label{IISW}
\end{eqnarray}
For LSS we have
\begin{eqnarray}
I_l^{\rm LSS}(k)= b_i \int^{\eta_0}_{\eta_{*}}\D z\,
n_{gi}(z) {\cal D}_{\Delta}[k; \eta(z)] j_l[k(\eta_0-\eta)],\label{ILSS}
\end{eqnarray}
where $b_i$ is the galaxy bias and $n_{gi}(z)$ is the redshift distribution of the galaxies normalized 
to $\int \D z\, n_{gi}(z) = 1$ for each redshift bin $i$.
In modified gravity, the growth rate is scale-dependent,
which would result in the scale-dependence of the bias.
However, as seen from Figs.~\ref{fig:D50.eps} and~\ref{fig:D5000.eps},
the growth rate is dependent only weakly on scales.
Therefore, we assume that the bias is scale independent for each redshift bin.


In the actual numerical calculation we use
the conventional scale-invariant primordial spectrum, $\Delta_{\cal R}^2(k)\simeq 2.4\times 10^{-9}$,
and the BBKS transfer function~\cite{bbks},
\begin{eqnarray}
T(k) = \frac{\ln(1+2.34 q)}{2.34 q}\left[1+
3.89 q+(16.1 q)^2+(5.46q)^3+(6.71q )^4
\right]^{-1/4},
\end{eqnarray}
where $q = k/\Gamma\,h$Mpc$^{-1}$ with
$\Gamma = \Omega_{\rm m0}h \exp(-\Omega_{\rm b0}-\sqrt{2h}\Omega_{\rm b0}/\Omega_{\rm m0})$~\cite{sugiyamashape},
$\Omega_{\rm m0}=0.3$, $\Omega_{\rm b0}=0.04$, and $h=0.7$.

To make the numerical calculations simpler by reducing the number of integrals,
we use the approximate formula~\cite{Limber1, Kaiser1, Kaiser2, Limber2}:
\begin{eqnarray}
\int F(x) j_l(k x)\D x\simeq \frac{\sqrt{\pi}}{2k}
\frac{\Gamma(l/2+1/2)}{\Gamma(l/2+1)}F[(l+1/2)/k].\label{lmb}
\end{eqnarray}
The idea that underlies the derivation of the above formula is that
the spherical Bessel function $j_l(x)$ monotonically grows from $0$ at $x=0$ to $x\simeq l+1/2$
and then starts to oscillate rapidly.
For $l\gg1$ we have $l+1/2\simeq l$ and $\Gamma(l/2+1/2)/\Gamma(l/2+1)\simeq\sqrt{2/l}$,
and Eq.~(\ref{lmb}) results in the Limber approximation often used in the literature.
The usual Limber approximation is accurate for $l\gtrsim 10$,
but the approximation at lower $l$ is improved by employing the formula~(\ref{lmb}). 
Indeed, the approximate results obtained from Eq.~(\ref{lmb})
are remarkably accurate even for small $l$,
as is confirmed by comparing them with
the exact numerical integration of $I^X_l(k)$.

\begin{figure}[t]
  \begin{center}
    \includegraphics[keepaspectratio=true,height=100mm]{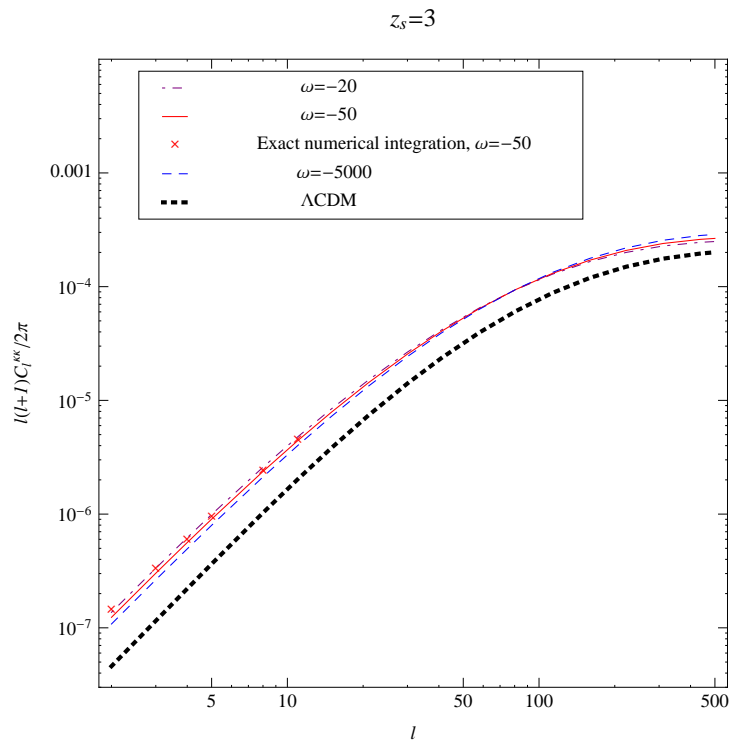}
  \end{center}
  \caption{Convergence power spectra for the Galileon models with $\omega=-20, -50, -5000$, and the $\Lambda$CDM model.
  The source location is given by $z_s=3$.}%
  \label{fig:wl3.eps}
\end{figure}

First, we compute the weak lensing convergence power spectrum defined by
\begin{eqnarray}
C_l^{\kappa\kappa} = \frac{l^2(l+1)^2}{4}C_l^{{\rm WL\,WL}},
\end{eqnarray}
which is shown
in Fig.~\ref{fig:wl3.eps}.
Here we have assumed that the source distribution is given simply by 
$n(\eta) = \delta\left(\eta-\eta(z_s)\right)$ with the source location $z_s=3$.
Although the background evolution is almost the same,
the convergence power spectrum in Galileon cosmology is enhanced compared to
that in the $\Lambda$CDM model, as is expected from enhanced $\Omega$ obtained in Sec.~\ref{num_calc}.
It is also found that the convergence power spectrum is
insensitive to the Brans-Dicke parameter $\omega$.
As has been done in many other studies~\cite{uzan, wlm1,wlm2,wlm3,wlm4,wlm5,wlm6,wlm7,wlm8,wlm9,Zhao}
(see, however, Ref.~\cite{bbk}),
the present analysis ignores
the non-linear growth of structure and non-linear recovery of GR,
which would be important for probing modified gravity with ongoing and future lensing surveys.
The non-linear effect will be more significant at higher multipoles ($l\gtrsim 200$).

\begin{figure}[t]
  \begin{center}
    \includegraphics[keepaspectratio=true,height=100mm]{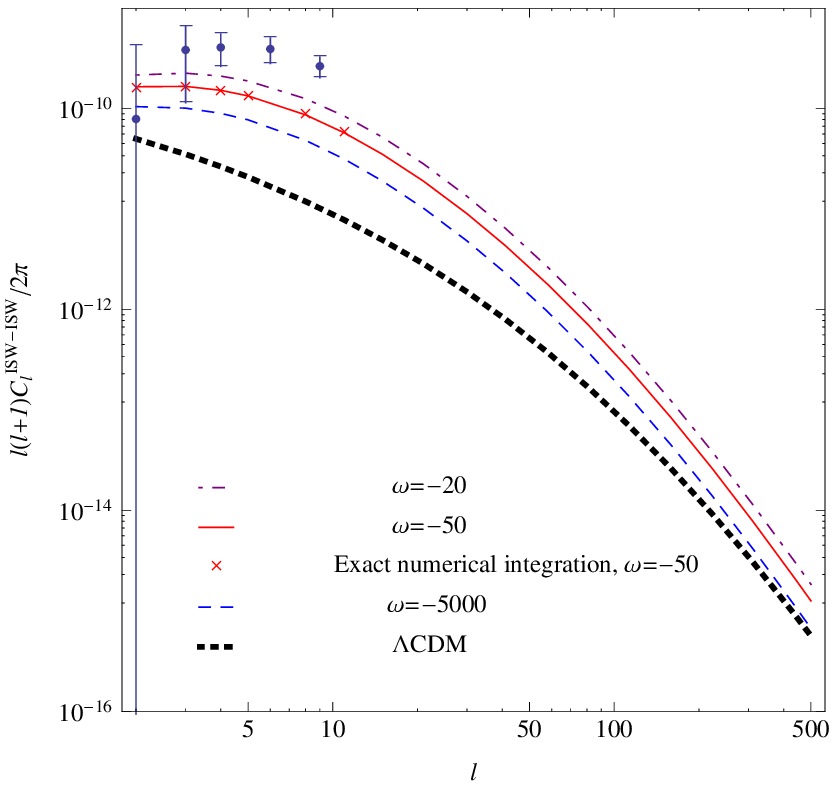}
  \end{center}
  \caption{ISW power spectra for the Galileon models with $\omega=-20, -50, -5000$, and the $\Lambda$CDM model,
  obtained by using
  the approximation formula~(\ref{lmb}). To check the accuracy of the approximation,
  the result obtained from the exact numerical integration of Eq.~(\ref{IISW}) is plotted by crosses.
  WMAP 5-year data for the total CMB angular power spectrum is also plotted with error bars.}%
  \label{fig:isw.eps}
\end{figure}

Next, let us discuss the ISW effect in Galileon cosmology.
The ISW effect is an additional cosmic microwave background (CMB) temperature anisotropy on large scales
caused by
the imbalance between blueshifts
experienced by the CMB photons falling into
the potential well and redshifts imprinted as they climb out.
The effect signals a net change of the gravitational potential along the path of the photons
due to dark energy or the modified gravity effect that mimics it.
Therefore, it is one of the most useful probes for dark energy and modified
gravity (see, e.g., Refs.~\cite{isw1,isw2,isw3,isw4,iswm, isw-dgp} for
the ISW effect in modified gravity models).
We show the ISW power spectra in Galileon cosmology in Fig.~\ref{fig:isw.eps}.
The ISW power spectrum is larger in Galileon cosmology than in the $\Lambda$CDM model
because of the larger time variation of the metric potential $\Omega$ in Galileon cosmology.
It can be seen that the amplitude is larger for smaller $|\omega|$.
Note, however, that whether $\Omega$ typically grows or decays is not evident
from the ISW auto-correlation.
(Growing $\Omega$ decreases the CMB temperature in Galileon cosmology,
while decaying $\Omega$ increases the temperature in the $\Lambda$CDM model.)

\begin{figure}[t]
  \begin{center}
    \includegraphics[keepaspectratio=true,height=100mm]{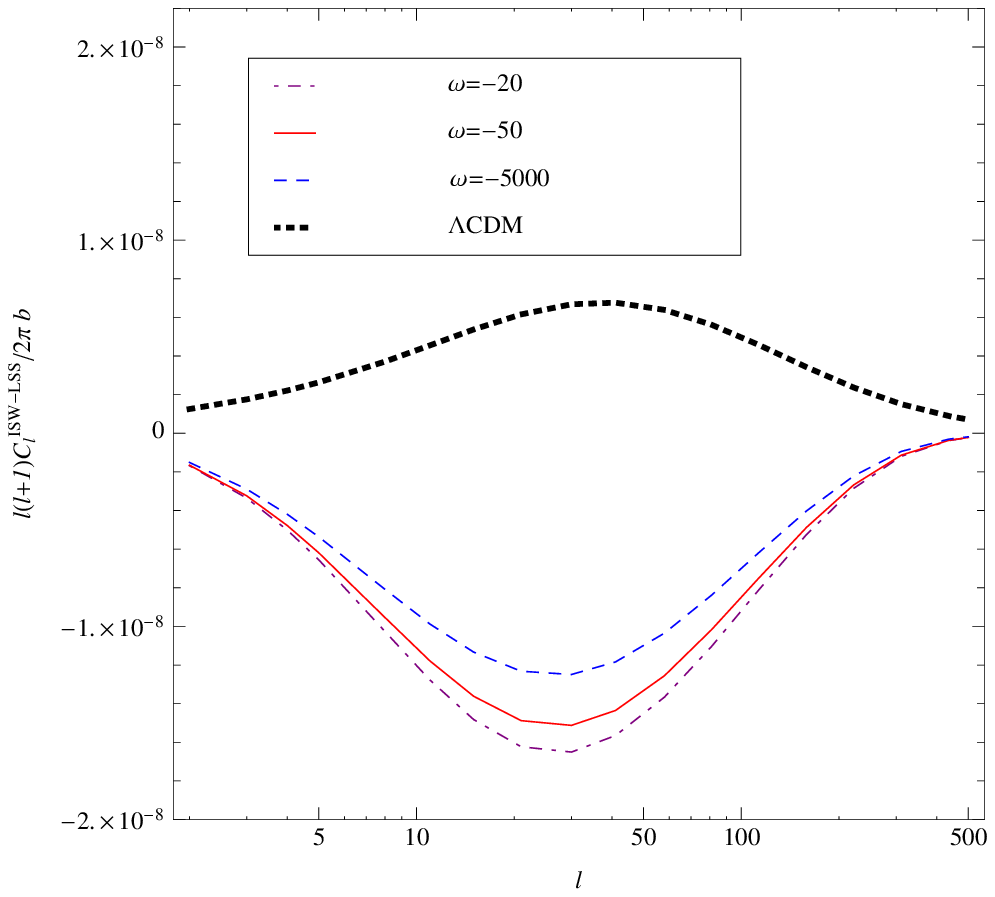}
  \end{center}
  \caption{Cross-correlation between ISW and LSS
  for the Galileon models with $\omega=-20, -50, -5000$, and the $\Lambda$CDM model.
  }%
  \label{fig:isw-g.eps}
\end{figure}

Unfortunately, the ISW effect is typically much smaller than the primary CMB temperature anisotropies and
it is only significant on the largest scales where cosmic variance is dominant. For this reason, 
it is difficult in practice to extract the ISW effect using CMB observations alone.
However, we can measure it by correlating CMB with LSS~\cite{CT}.
The ISW-LSS cross-correlation is shown in Fig.~\ref{fig:isw-g.eps},
where 
for definiteness, we assume that the galaxy distribution is given by
$n_{gi}(z) =1.5 \left(z^2/\bar z^3\right) \exp[-(z/\bar z)^{1.5}]$ with $\bar z=1$.
We find anti-correlations in Galileon models as a consequence of growing $\Omega$ at late times,
while the cross-correlation in the $\Lambda$CDM is positive due to the decaying potential.
The Galileon model with smaller $|\omega|$ shows a larger anti-correlation signal.
Negative ISW-LSS cross-correlations are also found in the normal branch DGP model~\cite{isw-dgp}.
The cross-correlation between CMB and LSS
has been evaluated using WMAP data and various different surveys of LSS,
and the detection of the positive cross-correlation has been 
suggested~\cite{isw,obs1,obs2,obs3,obs4,obs5,obs6,obs7,obs8,obs9,obs10}.
Therefore, the Galileon model is most strongly constrained from
the ISW-LSS cross-correlation.

\section{Conclusions}

In this paper, we have studied cosmological perturbations in
Galileon-type modified gravity~\cite{G1,G2,G3,SK}, i.e.,
scalar-tensor gravity with a non-linear derivative interaction of the form
$f(\phi)\Box\phi(\nabla\phi)^2$.
The gravitational theory is inspired by the effective theory in the DGP braneworld
and shares several interesting properties with the DGP model.
First of all,
the Galileon theory admits a self-accelerating solution without ghost-like instabilities,
and hence it can be an alternative to mysterious dark energy.
Moreover, the non-linear derivative interaction suppresses the scalar-mediated force
sufficiently near the source, evading solar-system and laboratory tests.
This non-linear screening mechanism (the Vainshtein effect) works also in a cosmological setting,
so that early-time standard cosmology can safely be restored.
To address the (super-)horizon scale evolution of metric and density fluctuations,
we have developed a full cosmological perturbation theory in Galileon-type models.
It was shown that the perturbation evolution is the same as
the $\Lambda$CDM case at early times, provided that the initial condition
for the Galileon fluctuation $\varphi:=\delta\phi/\phi$ is fine-tuned:
$\varphi_i=0=\dot\varphi_i$.
For generic initial conditions we found a super-horizon growing solution in the metric and Galileon perturbations,
implying that the homogeneous background we are considering is unstable.
Even if the initial condition is tuned to remove this growing solution,
the comoving density perturbation shows an unusual behavior
on super-horizon scales unless $\varphi=0$
at early times.
We thus argue that the most safe initial condition is to choose $\varphi_i=0=\dot\varphi_i$
at an early stage in the matter-dominated era.

We have focused on the evolution of
the particular combination of the metric potentials, $\Omega:=(\Psi+\Phi)/2$,
which determines weak lensing and the ISW effect and hence can help to constrain the Galileon model.
In the $\Lambda$CDM model, $\Omega\, (=\Psi=\Phi)$ decays when $\Lambda$ begins to dominate the universe.
In contrast to this, it was found that
in the Galileon model $\Omega$ grows rather than decays when
the effect of the Galileon field becomes large and the universe begins to accelerate.
A similar behavior can be found in the normal branch of the DGP model. 
We have computed explicitly the weak lensing convergence power spectrum
to show that it is enhanced relative to the $\Lambda$CDM case.
The resultant power spectrum was insensitive to the Brans-Dicke parameter $\omega$. 
Note that the effect of the non-linear
growth of structure must be taken into account in order to predict weak lensing signals correctly.
The non-linear effect becomes larger at higher multipoles.
We have also demonstrated that
the cross-correlation of the ISW effect and LSS shows an anticorrelation in Galileon cosmology,
which will be the most notable signature of the model
compared to the positive ISW-LSS cross-correlation in the $\Lambda$CDM model.
The cross-correlation was negatively larger for smaller $|\omega|$.
Since positive ISW-LSS cross-correlations 
have been suggested in the cross-correlation studies of
WMAP data with different LSS surveys, 
we conclude that this can provide the most stringent constraint 
on Galileon cosmology.

In this paper we have considered the specific form of the non-linear derivative interaction
$\sim f(\phi)\Box\phi(\nabla\phi)^2$ (mainly) with $f(\phi)=r_c^2/\phi^2$.
It would be interesting to explore the effects of higher order interaction terms~\cite{G1,G2}
on the background and perturbation evolution and on the ghost and stability issues.

\section*{Acknowledgements}
We thank Naoshi Sugiyama for useful comments.
TK is supported by the JSPS under Contract No. 19-4199.
HT is supported by the Belgian Federal Office for Scientific,
Technical and Cultural Affairs through the Interuniversity Attraction Pole P6/11.

\appendix
\section{Linearized field equations}

The $(00)$ and $(0i)$ components of the
perturbed Einstein equations are
\begin{eqnarray}
6H\left(\dot\Psi+H\Phi\right)-2\frac{\nabla^2}{a^2}\Psi
+\frac{\delta\rho}{2\phi}-\frac{\rho}{2\phi}\varphi
&=&3H\dot\varphi-6H I\Phi-3I\dot\Psi
-\frac{\nabla^2}{a^2}\varphi- \omega I\left(\dot\varphi- I\Phi\right)
\nonumber\\&&\quad + \phi^2 fI^2\biggl\{3I\dot\Psi
+2\left(6H-\alpha_1 I\right)I\Phi
-\left(9H-2\alpha_1 I\right)\dot\varphi
\nonumber\\&&\qquad
+\frac{\nabla^2}{a^2}\varphi+
 \left[-(2+\alpha_1)\left(3H-\frac{\alpha_1}{2} I\right)+\frac{\alpha_2}{2} I
\right]I\varphi
\biggr\},
\\
-2\left(\dot\Psi+H\Phi\right)
-\frac{\delta q}{2\phi}&=&-\left(\dot\varphi- I\Phi\right)
+H\varphi- (1+\omega)I\varphi
\nonumber\\&&\quad
+ \phi^2 fI^2\left\{
-I\Phi+ \dot\varphi
+\left[(1+\alpha_1)I-3H \right] \varphi
\right\},
\end{eqnarray}
and the trace part of the $(ij)$ equations is
\begin{eqnarray}
&&2\left[\ddot\Psi+3H\dot\Psi+H\dot\Phi+\left(3H^2+2\dot H\right)\Phi\right]
-\frac{\delta p}{2\phi}+\frac{p}{2\phi}\varphi
\nonumber\\&&\quad
=-2I\dot\Psi- \left[I\dot\Phi+2\left(\dot I+ I^2+2HI\right)\Phi\right]
+\ddot\varphi+2(H+ I)\dot\varphi
+ \omega I\left(\dot\varphi- I\Phi\right)
\nonumber\\&&\qquad\quad
- \phi^2 fI\biggl\{
- I^2\dot\Phi-2\left[2\dot I+(2+\alpha_1)I^2\right]I\Phi
\nonumber\\&&\qquad\qquad
+I\ddot\varphi+2\left[\dot I+ (2+\alpha_1)I^2\right]\dot\varphi
+ \left[(2+\alpha_1)\dot I+\frac{(2+\alpha_1)^2}{2}I^2+\frac{\alpha_2}{2}I^2\right]I\varphi
\biggr\}.
\end{eqnarray}
The traceless part of the $(ij)$ equations simply reads $\Psi-\Phi=\varphi$.
The equation of motion for $\phi$ is
\begin{eqnarray}
&&2\omega\left[
\ddot\varphi+(3H+ I)\dot\varphi-\frac{\nabla^2}{a^2}\varphi
- \left(2\dot I+I^2+6HI\right)\Phi- I\left(\dot\Phi+3\dot\Psi\right)
\right]
\nonumber\\&&\quad
+6\left[\ddot\Psi+4H\dot\Psi+H\dot\Phi
+\left(4H^2+2\dot H\right)\Phi\right]-2\frac{\nabla^2}{a^2}(2\Psi-\Phi)
\nonumber\\&&\qquad
-\phi^2 f \biggl\{
-4I(3H- \alpha_1 I)\ddot\varphi
+4\left[-3\left(H\dot I+\dot H I+3HI^2+3H^2I\right)
+2 \alpha_1 I\dot I+ \left(\alpha_1^2+3\alpha_1+\alpha_2\right)I^3\right]\dot\varphi
\nonumber\\&&
+\varpi\varphi
+4\left(\dot I+ I^2+2HI\right)\frac{\nabla^2}{a^2}\varphi
+6\left[I^2\ddot\Psi+2\left(I\dot I+ I^3+3HI^2\right)\dot\Psi\right]
+2I^2\frac{\nabla^2}{a^2}\Phi
\nonumber\\&&\quad
+2\left(9H-2\alpha_1 I\right)I^2\dot\Phi
+4\left[
6I\left(2H\dot I+\dot H I+2HI^2+3H^2I\right)-4\alpha_1I^2\dot I
- \left(\alpha_1^2+3\alpha_1+\alpha_2\right)I^4
\right]\Phi
\biggr\}=0,
\end{eqnarray}
where
\begin{eqnarray}
\varpi&:=&-6(2+\alpha_1)\left(2H\dot I+\dot HI+2HI^2+3H^2I\right)I
+4\left[\alpha_1(2+\alpha_1)+\alpha_2\right]I^2\dot I
\nonumber\\&&
+ \left[
\alpha_3+\alpha_2(5+3\alpha_1)+\alpha_1(3+\alpha_1)(2+\alpha_1)
\right]I^4.
\end{eqnarray}



\end{document}